\documentclass[preprint]{aastex}
\begin{document}
\pagestyle{empty}
\begin{center} \large{\bf H$\alpha$ emission line morphologies in Markarian starburst
galaxies\\}
\end{center}
\noindent
\centerline{A. Chitre$^{1}$ and U. C. Joshi}
\noindent
\centerline{Physical Research Laboratory, Navrangpura, Ahmedabad 380 009, India}
\centerline{$^{1}$present address: Indian Institute of Science, Bangalore}
\vskip 4in
\noindent
Running title : H$\alpha$ morphology in Markarian starburst galaxies\\




\newpage
{\bf Abstract} We present broad band $R$ and narrow band H$\alpha$ emission line images of a
sample of
optically selected starburst galaxies from the Markarian lists.
The emission line morphology is studied and global properties like luminosities,
equivalent widths and star formation rates are derived.
The radial distribution of H$\alpha$ flux
and the EW are determined using concentric aperture
photometry on the emission line and the continuum images. H$\alpha$ flux is
generally found to peak in the nuclear region and fall off outwards. The EW
is found to peak off-center in most of the cases implying that though the intensity
of emission is maximum at the nucleus, the star formation activity relative
to the underlying continuum often peaks away from the center in Markarian
starburst galaxies.\\
{\em keywords:} galaxies: starburst; galaxies: photometry

\section{Introduction}

Emission line fluxes of the Balmer series provide a measure of the integrated luminosity of galaxies
short ward of the Lyman limit and can be used to provide a direct measurement
of the star formation rate (SFR) for hot, massive stars that dominate the
ionizing continuum. The H$\alpha$ emission from giant HII regions in
galaxies is a good tracer
of OB star formation. The presence of H$\alpha$ indicates that at least one
massive star has formed there recently. Hence, H$\alpha$ images are ideal for
tracing the spatial distribution of hot, young stars in star forming regions
of normal
 and starburst galaxies.
Extensive H$\alpha$ surveys of samples of
normal galaxies were carried out by Kennicutt \& Kent (1983). They
found that the integrated emission of a galaxy is strongly correlated with
its Hubble type and colour. They also inferred that the variation of emission
among galaxies of a given type was due to real dispersion in star formation
activity.
Several H$\alpha$ surveys of large samples of spirals using large apertures
(Romanishin 1990) and CCD images (Garc\'{i}a-Barreto et al. 1996; Pogge 1989) were made. From a CCD H$\alpha$ survey, Gonzalez Delgado \& Perez (1997a, 1997b) report considerable circumnuclear emission in a large sample of active galaxies. Pogge \& Eskeridge (1993) conducted
an H$\alpha$ imaging survey of a sample of neutral-gas rich S0 galaxies.
Starburst galaxies and violently interacting systems show enhanced levels of star formation. The nuclear as well as the global properties of violently interacting galaxies
were studied in detail by Bushouse (1986).
Lehnert \& Heckman (1996), hereafter LH96;  studied a sample of IR selected edge-on
starburst galaxies using H$\alpha$ images.
In this paper, we present the H$\alpha$ images along with the $R$ band images of a sample of Markarian starburst galaxies and discuss the properties of the line emission. This paper is the second in a series of papers aimed at studying
Markarian starburst galaxies in $B,V,R,I$ and H$\alpha$. 
The optical $B,V,R,I$ properties of a subset of the sample of Markarian starburst galaxies
can be found in Chitre \& Joshi (1999), hereafter, PaperI.
\section{Observations and data reduction}
The sample of galaxies was selected from the Markarian lists (Markarian,
Lipovetskii \& Stepanian 1979; and references therein) with the following criteria:
Galaxies designated as starburst galaxies in the catalogue from CDS with m$_v$ brighter than
$14^m.5$ and angular sizes greater than 30\arcsec\ were
selected. No consideration was made for any particular morphological type
and the sample is unbiased towards the morphology of the galaxy. In this paper,
we present results on 15 of the sample objects. All morphological
types are represented in the present sample. Table 1 lists
the global properties
of the galaxies presented.

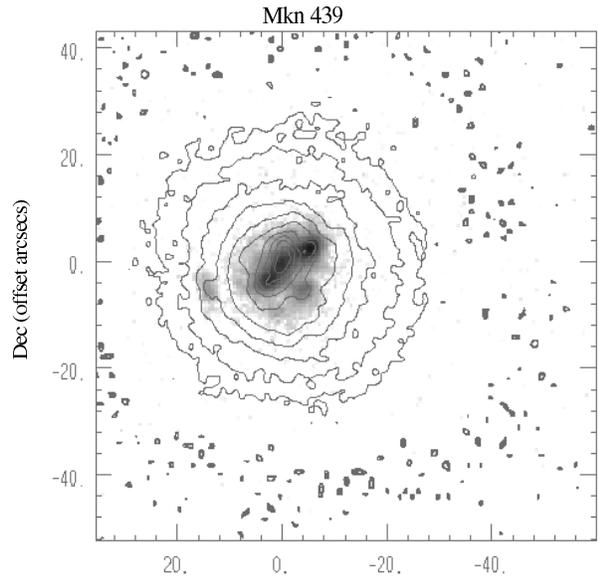
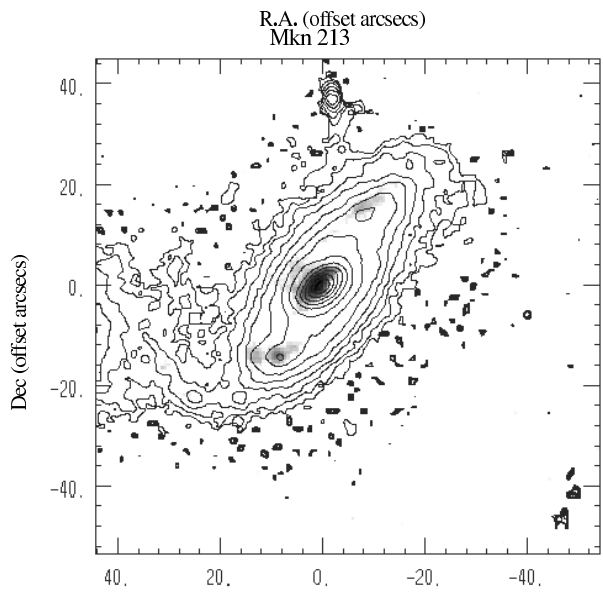
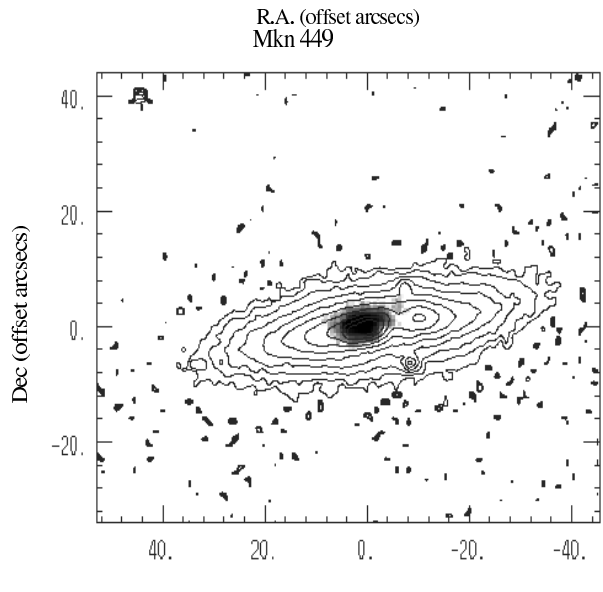
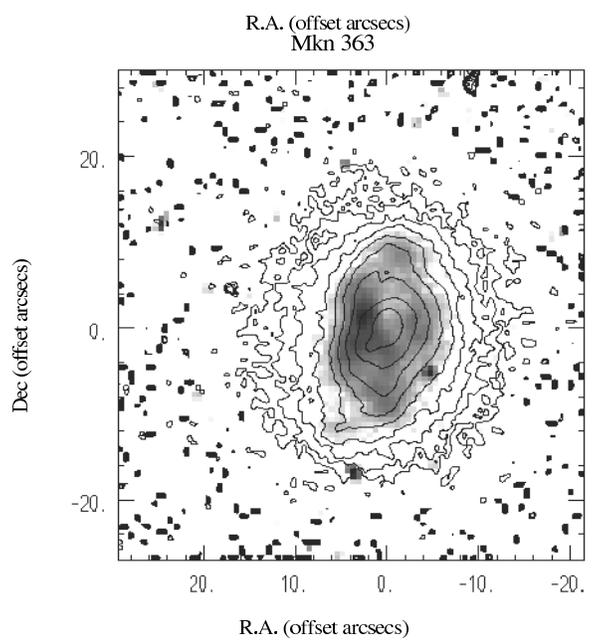
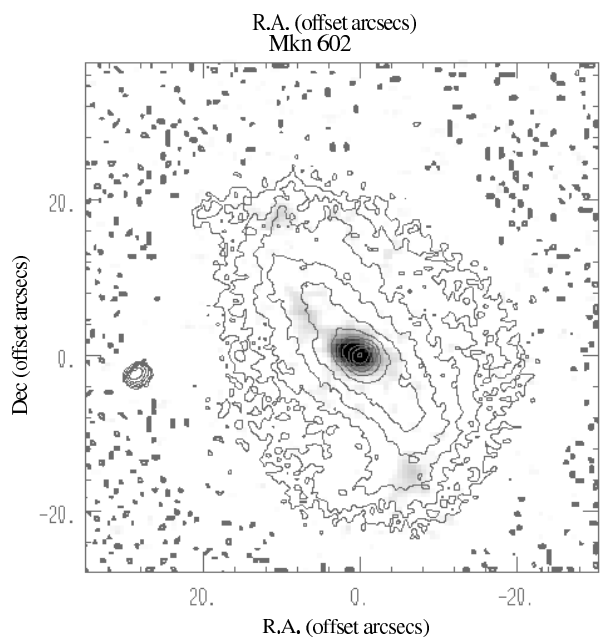
\begin{figure}[ht]
\vskip 5mm
\caption{$R$ band continuum isophotal contours superimposed on the H$\alpha$ + 
[NII] emission line gray scale images. The gray scale has been chosen for best
contrast. North is at the top and East is to the left. The $R$ band contours are
plotted on the magnitude scale at an interval of 0.5$^m$ with the peak contours 
at 16.5 mag arcsec$^{-2}$ for Mkn 213, Mkn 603 and Mkn 1379; 17.0 mag arcsec$^{-2
}$ for Mkn 439, Mkn 602 and Mkn 1194; 17.5 mag arcsec$^{-2}$ for Mkn 449, Mkn
708 and Mkn 781; 18.0 mag arcsec$^{-2}$ for Mkn 14, Mkn 363, Mkn 743, Mkn 1002 and
 Mkn 1308 and 18.5 mag arcsec$^{-2}$ for Mkn 1134.}
\label{contour}
\end{figure}

The observations were carried out at the Cassegrain focus of the 1.2m
telescope at Gurushikhar, Mt.Abu. The f-ratio of this telescope is f/13.
A thinned, back illuminated Tektronix 1024 $\times$ 1024 pixels CCD chip 
providing a field of view (5.2$\arcmin \times $5.2$\arcmin $) was
used for imaging the galaxies. On chip binning of 2 $\times$ 2 was employed before recording the
images to increase the signal-to-noise ratio. The final resolution was
0.63\arcsec\ /pixel which is sufficient to sample the point spread function (PSF) appropriately. The PSF values during each galaxy observation are
given in Table 3.
Images were obtained through broadband
Cousin's $R$ and narrow band H$\alpha$ filters. The sample of galaxies was
observed using appropriately red shifted filters. The response curves of the
narrow band filters used are given in Fig. \ref{nbf}. Table 2 gives the filter specifications viz. the mean wavelength ($\lambda_0$) the peak transmission and the effective width (the width of a rectangle with the same area as that under the response curve and height equal to the peak transmission; W) of each filter.
About 3-4 exposures were taken in each filter. The exposure times are also given in
Table 3. The data was
reduced using IRAF \footnote{IRAF is distributed by National Optical Astronomy
Observatories, which is operated by the Association of Universities Inc.
(AURA) under cooperative agreement with the National Science Foundation,
USA.} and standard reduction procedures viz. bias subtraction, flat fielding,
cosmic ray correction were performed on the raw images. Airmass corrections
were applied using the extinction coefficients obtained from the standard stars
which were observed on the same nights together with the narrow band and the
broad band images. The galaxy images in both $R$ band and narrow band H$\alpha$ were
shifted and co-added to improve the signal-to-noise ratio. The sky background
was computed from the mode of the histogram of the image since the field of view
was large enough as compared to the size of the galaxy and the histograms were
sky dominated. The galaxy images were subsequently sky subtracted and scaled for 1 second exposure. Figure \ref{contour} displays the $R$ band continuum isophotal contours superimposed on the H$\alpha$ emission line images.
\begin{figure}[ht]
\caption{Normalized response curves of the narrow band filters.}
\label{nbf}
\end{figure}

\section{Generation and calibration of pure emission line images}
The images obtained as discussed above were further reduced in the following way to obtain the pure
emission line images. The FWHM of the stars in each set of $R$ and H$\alpha$
frames was examined and the better PSF was degraded to match the PSFs in $R$
and the narrow band. The next step involved determining the scale factor
between the broad band and the narrow band image. For this, all the 
foreground stars in each frame were used. Using synthetic aperture photometry, the
counts within a fixed aperture were derived for stars in both the filters and their ratio was taken.
The average scaling factor $\epsilon$, between the $R$ band and H$\alpha$ was
thus determined for each galaxy. The $R$ band image was then scaled by the
corresponding factor and subtracted from the narrow band images to yield the
pure H$\alpha$+[NII] emission image.

The zero point $R_0$ for the continuum $R$ band was determined using 
\begin{equation}
R=R_0-2.5\, {\rm log}N
\end{equation}
where $N$ is the number of counts and $R$ is the corresponding instrumental
magnitude. We used our broad band data (Chitre 1999) to determine the zero point for each observing night.
The $H\alpha$ fluxes were determined using the equations explained in detail by Macchetto et al. (1996). These values were then corrected for galactic absorption using
$E(B-V)$ from Burstein \& Heiles (1984) and the Whitford (1958) form of the extinction curve as:
\begin{equation}
f_{H\alpha corr} =
f_{H\alpha obs} \times {\rm dex}(0.969 \times E(B-V))
\label{ext}
\end{equation}
We have not applied
corrections for the internal extinction within the galaxy.
The corrected fluxes were converted to luminosities using the distances given in Table 1.
 The global equivalent widths of the galaxies were estimated by dividing the
corrected H$\alpha$ flux in the emitting region of the galaxy by the
underlying $R$ band continuum flux within D$_{25}$. The underlying $R$ band continuum was determined by first subtracting the emission from the broad band image and then scaling it to the continuum in the narrow band. The global EW and 
the $B-V$ colours of the galaxy are plotted in Fig. \ref{ewbv}. We do not find a
strong correlation between the
global EW and 
the $B-V$ colours of the galaxy. There is a large scatter in EW around $B-V$ of 0.6. 
All the derived quantities are presented in Table 4.

\begin{figure}
\caption{H$\alpha$ equivalent widths (present study) vs. B-V colours (Chitre 1999)} 
\label{ewbv}
\end{figure}

\section{Discussion}
\subsection{Spatial distribution of star formation}
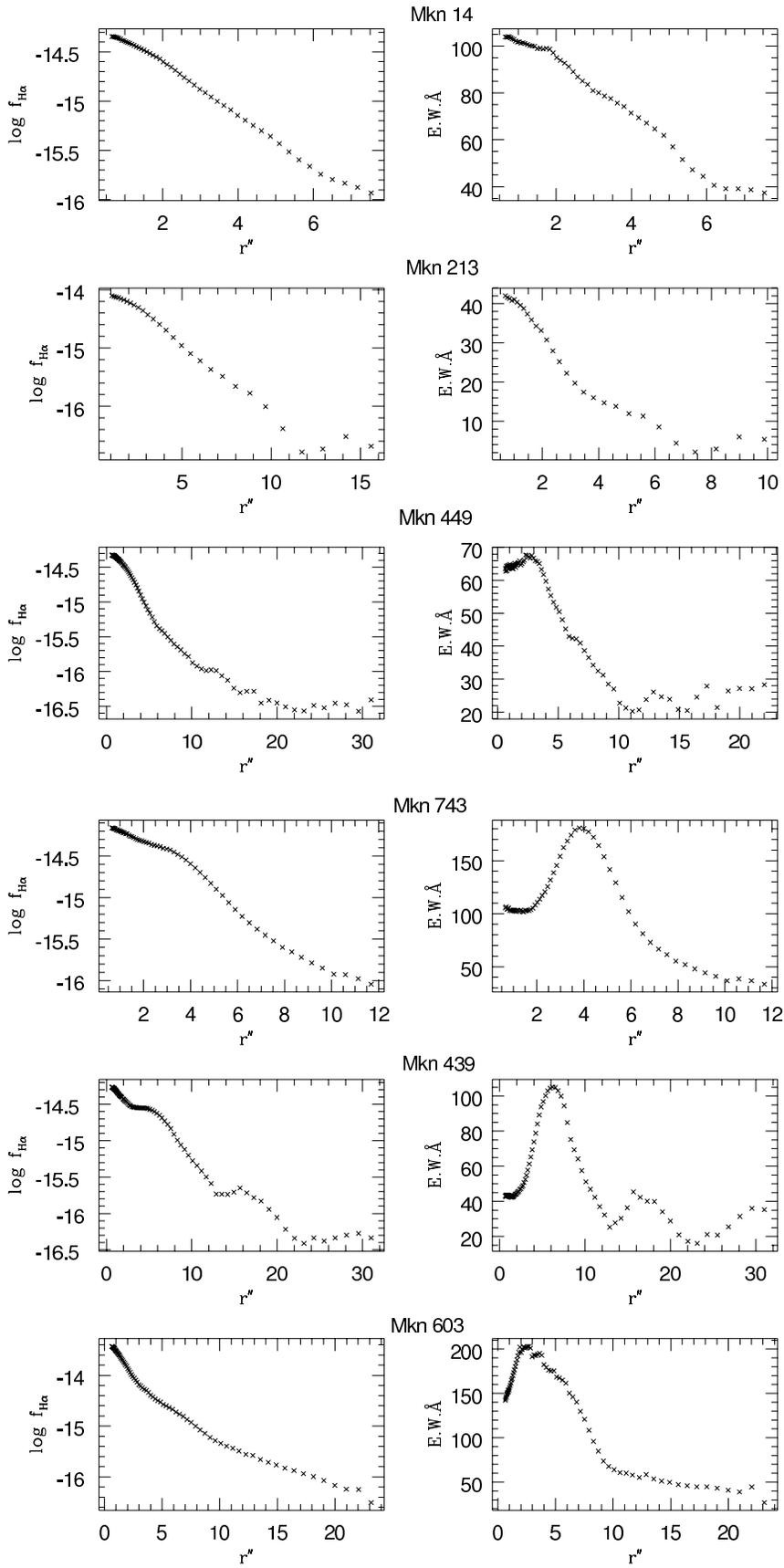
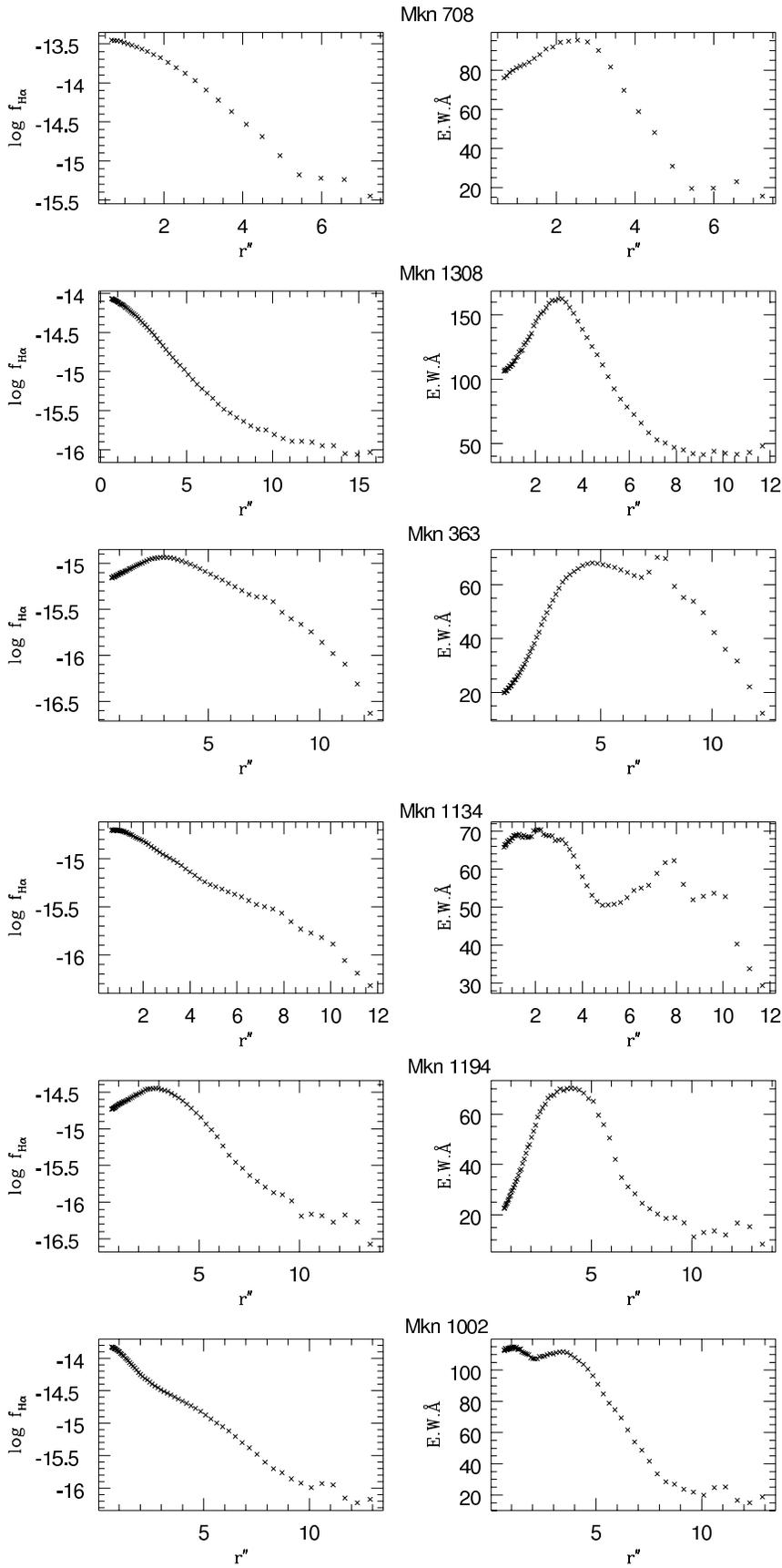
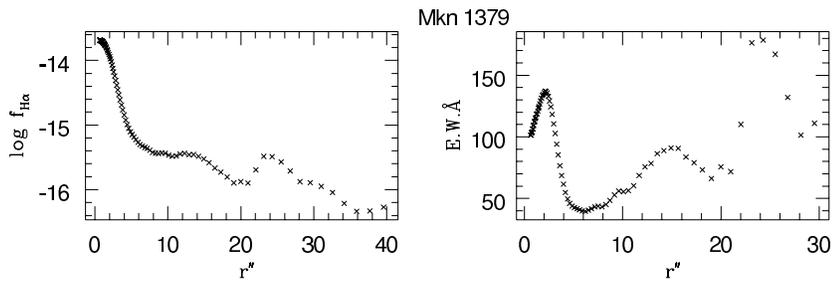
\begin{figure*}
\vskip 5mm
\caption{The radial variation of the logarithm of the H$\alpha$ flux (erg  s$^{-1}$ cm $^{-2}$) and the EW (\AA)}
\nonumber
\label{rew}
\end{figure*}

The large-scale distribution of star formation can be studied in various ways.
The simplest qualitative method is to visually examine the distribution of the
emission line regions.
On the basis of the emission line morphologies, the sample can be divided into
the following four sub classes.
\begin{enumerate}
\item Galaxies showing H$\alpha$ emission in the central regions
only. Mkn 14, Mkn 449 and Mkn 1308 belong to this sub class.
\item Galaxies showing extended H$\alpha$ emission or galaxies
with extranuclear emission in addition to nuclear star formation.
This includes most of the
spirals which show emission from either one or both the ends of the
bar or along the bar. Mkn 213, Mkn 602 are examples of this sub class.
\item Galaxies showing global massive star formation. Line emission
is observed throughout the body of the galaxy in Mkn 363, Mkn 603 and Mkn 1134,
indicating that these galaxies are undergoing a global starburst.
The fact that massive star formation is not always confined to the
nuclear region alone is best illustrated in these cases. Their strong
stellar continuum at longer wavelengths ($R$ band) points to the fact
that these are not young systems experiencing their first phase of star
formation, but are old systems with a younger burst of star formation.
\item This sub class includes starburst galaxies with peculiar emission line
morphologies. Mkn 439 and Mkn 1194 are examples of such systems. Mkn 439 shows
extended H$\alpha$ emission along a bar which is misaligned
with the optical continuum
isophotes (Chitre, Joshi \& Ganesh 1999), while Mkn 1194 shows
very faint emission from the central region but clearly shows a
circumnuclear ring of intense massive star formation.
\end{enumerate}
However, to study the emission line distribution in a quantitative manner
and derive useful information from the results, synthetic aperture 
photometry was performed on the continuum subtracted H$\alpha$ images
to derive the radial flux distribution. Concentric apertures were
used for this purpose. The apertures were centered on the nucleus, the position
of which was determined from the $R$ band continuum images. This convention
was maintained irrespective of whether or not the H$\alpha$ peak coincides with
the $R$ band peak. Circular apertures were considered up to the radius where the
signal falls to the 3$\sigma$ level of the background value. The radial
distribution of the H$\alpha$ flux thus obtained is plotted in Fig. \ref{rew}.
To study the regions of current star formation relative to the underlying
continuum contributed by older stars, the equivalent width (EW) was computed.
This was done by dividing the H$\alpha$ flux by the underlying stellar continuum
flux at the same location estimated from the emission subtracted, scaled continuum images. It is a good measure for
comparing relative levels of star formation from one galaxy to another.
The EW provides a measure of the current star formation rate relative to the recent past
star formation rate, since the line emission is due to massive stars with ages
$<$ 10$^7$ years while the stellar continuum at these wavelengths is due to G and
K giants which are typically a few billion years old (Huchra 1977,
Kennicutt 1983). A similar approach was used by Bushouse (1986)
to derive pseudo equivalent widths for a sample of interacting galaxies.

An inspection of Fig. \ref{rew} reveals that most of the galaxies except
Mkn 363 and Mkn 1194 show a peak in the flux at the center and a nearly
exponentially falling behaviour outwards irrespective of their morphology.
In some cases like Mkn 1308, the profile flattens out in
the outer regions. In case of Mkn 363 and Mkn 1194, the peak emission is
off-centered from the nucleus. This is clearly seen in the gray scales
images (Fig. \ref{contour}) also. In general, galaxies with peculiar
emission line morphologies or global emission show considerable
deviations from the exponential profiles eg. Mkn 1134 and Mkn 439. However,
the EW profiles show startling differences in their behaviour. The first two
panels in Fig. \ref{rew} show galaxies with EWs peaking at the center
and falling outwards. The next few panels have EWs peaking away from
the center. The interesting point to be noted in these cases is that the
EW peaks away from the center even in those cases where the peak H$\alpha$
flux is maximum at the center. This indicates that though the flux levels
are maximum at the center in some of these galaxies, the current star
formation rate relative to the recent past star formation is maximum away
from the center. The peak occurs in the inner half a kpc in most of these galaxies except in Mkn 363, Mkn 1134, Mkn 1194 and Mkn 1379 all of which show considerable extended extranuclear star formation. The peak is sharp in most cases except in Mkn 363
where it is broad and flattened. This is probably a result of the global
star formation seen in Mkn 363.
The global EW listed in Table 3 was compared with the EW
in the central 1 kpc of each galaxy except for Mkn 602 and Mkn 781. These galaxies were excluded because the signal-to-noise ratio was not good enough to derive the radial profiles and only global values were computed for these two galaxies. It was found that the central nuclear EW was higher than
the global value in nearly all cases.  This suggests that the star formation is
more enhanced with respect to the underlying population in the central regions
of these starburst galaxies. A few cases like Mkn 363 and Mkn 1194
have the ratio of the nuclear EW to its global value less
than unity.
Both these galaxies show a considerable amount of extranuclear star
formation.

\subsection{Half-light radii and the relative concentration of star formation}
To compare the characteristic sizes of the star forming regions in the sample,
we derived the half-light radius in H$\alpha$. This parameter enables us
to define a model independent size of the emitting region especially while
inter comparing data from a morphologically dissimilar sample as ours. 
The half-light radius in H$\alpha$ ($r_{e,H\alpha}$) is defined as the radius
enclosing half of the total line emission. For deriving this, the radius at
which the counts in the image frame fall to 2$\sigma$ of the background value
is determined.  The total flux inside this aperture is computed.
A growth curve is constructed using concentric apertures of increasing
radii and the half-light radius is determined from this curve.
The values of $r_{e,H\alpha}$ derived in this manner are given in Column (7)
of Table 4. The mean value of log ($r_{e,H\alpha}$) derived from our sample is 2.88
in pc. The size of the emitting regions is directly proportional to
the absolute luminosity of the galaxy. This is illustrated in Fig. \ref{mag}
where the absolute blue luminosity M$_B$ is plotted against the half-light
radius in H$\alpha$, $r_{e,H\alpha}$. 

\begin{figure}
\caption{The absolute B magnitude as a function of the half-light radius in H$\alpha$}
\label{mag}
\end{figure}

To estimate the degree of concentration of the emission relative to the continuum, we calculated the ratio of the half-light radius in H$\alpha$ to
the half-light radius in $R$ ($r_{e,R}$). The values of $r_{e,R}$ were taken from Chitre (1999). For our sample, this ratio is less than unity for all cases indicating that the emission is more centrally concentrated as compared 
to the continuum light. To compare the properties of our sample of optically selected, uv-excess starburst galaxies with other IR selected starburst galaxy samples, we have done a comparative study of our sample with LH96 sample.
The LH96 sample was selected based on the following criteria:
It contained starburst galaxies which were edge on,
were IR bright ($S_{60\mu m}$ $>$ 5 Jy) and IR warm ($S_{60\mu m}$$/$$S_{100\mu m}$$>$0.5. Our sample galaxies had a range of inclinations with Mkn 439 being nearly face on and Mkn 1194 being nearly edge on. Our sample galaxies were selected only on the basis of their redshifts and apparent magnitudes and were found to have
$S_{60 \mu m}$ ranging from 0.25 Jy to 12.6 Jy with most galaxies having
the value less than 5 Jy. Comparing the results derived by us with the results
derived by LH96 (Table 5), we see that the average size of the emitting
regions in the LH96 sample is nearly 1.8 times larger than that in our sample.
The concentration of H$\alpha$ in our sample is 45$\%$ with respect to the continuum
while it is 60$\%$ for LH96. In other words, the line emission is more centrally concentrated in our sample of optically selected starburst galaxies
than in the LH96 sample of IR selected starburst galaxies. We also find that
on an average, LH96 galaxies are only 1.5 times brighter in H$\alpha$ but are
about 9 times brighter in FIR. One reason for this could be that the IR selected galaxies are much dustier and the star forming regions are hidden by large amounts of dust which could be responsible for the extinction of the line emission.

\begin{figure}
\caption{log($L_{IR}/L_{H\alpha}$) vs r$_{e,H\alpha}$/r$_{e,r}$ for our sample compared with two other samples.} 
\label{plot2}
\end{figure}
Fig. \ref{plot2} shows a comparative study of the behaviour of
log($L_{IR}/L_{H\alpha}$) versus r$_{e,H\alpha}$/r$_{e,r}$ for three
samples, namely our sample, LH96 IR selected starburst galaxies sample and Usui et al. (1998)
sample which consists of early-type spirals with log(L$_{FIR}$/L$_B$) higher than the average for early-type spirals.
Our sample objects have smaller ratios of both log (L$_{IR}/$L$_{H\alpha}$)
and $r_{e,H\alpha}/r_{e,R}$ as compared to LH96 while the values do not show much difference as compared to the Usui sample of early-type star forming galaxies. 
\section{Conclusions}
A study of a sample of Markarian starburst galaxies has shown that
\begin{enumerate}
\item The line emission peaks in the central region and falls nearly exponentially outwards in most of the cases.
\item Unlike the emission, the radial variation of the equivalent width
does not show an
uniform behaviour. This indicates that though the intensity of star
formation is maximum at the center, the relative level of star formation
with respect to the underlying continuum shows different trends. We do not find any clear evidence for a relation between the radial behaviour of the equivalent width and the morphological type of the galaxy.
\item The nuclear EW is greater than the global EW in most of the sample
galaxies.
\end{enumerate}
\acknowledgements
This work was supported by the Department of Space, Government of India.
The authors are thankful to Mr. Shashikiran Ganesh for helping with
observations and to Mr. A.B. Shah, Mr. N.M. Vadher
for technical support. This research has made use of the VizieR catalogue
access tool, CDS, Strasbourg, France.

\appendix
\section{Notes on individual galaxies:}
The emission line morphology of each galaxy is described below and
compared with the optical colour maps from PaperI or Chitre
(1999).

{\it    Mkn 14} : This S0 galaxy shows line emission in the central region.
The emission appears to be more concentrated towards the eastern side of the
nucleus (Fig. \ref{contour}).

{\it Mkn 213} : This spiral galaxy shows strong H$\alpha$ emission in the
central region fanning out towards the west and faint emission at the ends
of the bar. The EWs are low compared to other galaxies in the sample
indicating that the older stellar population is dominant in the inner region in
this galaxy.

{\it Mkn 363} : A peculiar emission morphology is seen in this galaxy.
H$\alpha$ emission is global but the morphology of the emission is disturbed and
 clumpy. The peak of emission off-centered from the nucleus.
A comparison with the optical colour maps shows that the bluest region in this
galaxy also lies 5\arcsec\ away from the nucleus.
The EW (Fig. \ref{rew}) also reaches its peak
value at this distance.

{\it Mkn 439} : This galaxy is the most peculiar galaxy in the sample.
The optical continuum contours and the emission line contours show completely
different morphologies. The peak emission in H$\alpha$ does not coincide with
the bluest region seen in the optical colour maps. The line emission is
along a bar misaligned with the optical contours of the galaxy (Chitre, Joshi
\& Ganesh
 1999). The EW peaks 6\arcsec\ away from the optical center of the galaxy.

{\it Mkn 449} : Strong H$\alpha$ emission is seen only in the central region in
this
galaxy.

{\it Mkn 602} : This barred spiral shows emission in the nuclear region and very
faint emission from the ends of the bar.

{\it Mkn 603} : Mkn 603 is a part of an interacting system of galaxies. The main
galaxy is an elliptical interacting with two small companions. This galaxy shows
extended line emission while its companions show global line emission.

{\it Mkn 708} : An elongated spiral galaxy with very broad ill-defined spiral arms, Mkn 708
shows line emission only in the central region. However, this galaxy get redder inwards unlike other Markarian starburst galaxies which get bluer inwards.
We do not find any clear indications for the
presence of dust from isophotal analysis of the optical continuum images (Chitre 1999).

{\it Mkn 743} : This double nucleus galaxy (Mazzarella \& Boroson 1993) shows peak H$\alpha$ emission in one
of its nucleus. Emission is seen in the inner region in this galaxy.

{\it Mkn 781} : The only flocculent spiral in the sample, this galaxy shows
emission in the nuclear region and as a couple of very faint spots in the arms.

{\it Mkn 1002} : Pogge \& Eskridge (1993) have reported H$\alpha$
emission in the
nuclear region as well as clumps of emission in the circumnuclear region
of this galaxy. Our studies show that the line emission does not follow
the structure seen in colour maps but just shows extended emission
in the central region.

{\it Mkn 1134} : Global star formation is seen in this small irregular galaxy
which is attached to the tip of the spiral arm of a larger spiral galaxy,
NGC 7753. The peak emission is coincident with the off-centered nucleus
of the galaxy.

{\it Mkn 1194} : Star formation is enhanced in a circumnuclear ring in this
galaxy. Two bright H$\alpha$ spots are seen along the ring to the north
and the south. The nucleus appears faint in emission. The ring lies at about
5\arcsec\ from the nucleus. The EW is also found to peak in this region.

{\it Mkn 1308} : This S0 galaxy has an elongated companion. Emission is seen in
the nuclear region. However, the EW peaks 3\arcsec\ from the optical center of
the galaxy.

{\it Mkn 1379} : Mkn 1379 forms a part of a system of interacting galaxies.
Strong H$\alpha$ emission is seen in the nuclear region and diffuse emission is
seen along the bar. Blobs of H$\alpha$ are seen at the ends of the bars and
in the companion galaxies. Global star formation is detected in the companions
lying to the east. We measured the EWs in these eastern companions to be as
high as 245\AA. The companion to the west has a EW of about 105\AA.
These values are much higher as compared to the peak EW seen in Mkn 1379
(Fig\ref{rew}).
\clearpage
\begin{table}
\caption{Global properties of the sample galaxies}
\begin{tabular}{llllllclll}
\tableline
\tableline
Mkn&NGC&UGC&Type&Dist.&S$_{60\mu m}$&S$_{60\mu m}$/S$_{100\mu m}$&log L$_{FIR}$&
B$_T$&log D$_{25}$\\
&&&CDS&Mpc&Jy&&erg s$^{-1}$&&RC3\\
(1)&(2)&(3)&(4)&(5)&(6)&(7)&(8)&(9)&(10)\\
\tableline
14&-&4242&S0?&44&0.25&0.23&42.69&14.67&0.73\\
213&4500&7667&SBa&43.6&3.89&0.59&43.67&13.18&1.21\\
363&694&1310&Scp&41.5&2.4&0.65&43.40&14.28&1.58\\
439&4369&7489&Sa&14.5&5.91&0.53&42.92&12.65&1.32\\
449&5014&8271&Sap&15.4&2.27&0.56&42.54&13.45&1.23\\
602&-&2460&SBbc&38.2&3.55&0.63&43.50&13.50&1.09\\
603&1222&-&E&35.8&12.6&0.82&43.96&13.24&1.04\\
708&2966&5181&SB&21.2&5.37&0.672&43.16&12.53&1.35\\
743&3773&6605&E0p&11.7&1.48&0.83&42.06&13.19&1.07\\
781&4779&8022&SBc&36.6&1.82&0.44&43.25&13.07&1.33\\
1002&632&1157&E1&43.3&4.85&0.76&43.72&13.60&1.19\\
1134&7752&12779&I0&72&4.76&0.48&44.23&14.75&0.92\\
1194&1819&3265&SB0&62.1&6.95&11.40&44.23&13.23&1.22\\
1308&-&6877&-&11.9&0.99&1.29&41.84&14.04&0.82\\
1379&5534&-&SBbc dbl&36.5&4.73&0.68&43.58&13.29&1.15\\
\hline
\end{tabular}
\vskip 5mm
\tablecomments
{
Cols. (1),(2) and (3).- The Mkn, NGC and UGC designations respectively of the ob
jects.
Col. (4).- The morphological type taken from the Markarian catalogue downloaded
from CDS.
Col. (5).- The adopted distance to the galaxy in Mpc.
Col. (6).- The 60 $\mu m$ flux density in Jy from IRAS.
Col. (7).- The ratio of the 60 $\mu m$ to the 100 $\mu m$ flux density.
Col. (8).- The logarithm of the FIR luminosity in ergs s$^{-1}$ calculated using
 the formula from Lonsdale et al. (1985).
Col. (9).- The total uncorrected B magnitude from Chitre (1999). This w
as derived by fitting an exponential disk to the outer part of the luminosity pr
ofile in the B band and extrapolating it to infinity to derive B$_T$.
Col. (10).- The log of the diameter at 25 mag arcsec$^{-2}$ in units of a tenth
of an arc minute from RC3.
}

\end{table}
\clearpage
\begin{table}
\caption{Narrow-band filter specifications}
\begin{tabular}{cccc}
\hline
\hline
Filter no.&$\lambda_0$&Peak trans.& W(eff.width)\\
          &(\AA)&(\%)&(\AA)\\
\hline
F1&6556&60&91\\
F2&6618&60&84\\
F3&6669&65&81\\
\hline
\end{tabular}
\end{table}

\clearpage

\begin{table}
\caption{Observational details}
\begin{tabular}{lllcl}
\hline
\hline
Mkn&$t_R$&$t_{H\alpha}$&Filter&PSF\\
&sec&sec&no.&$\arcsec$\\
\hline
14&480&2100&F2&3.0\\
213&660&2100&F2&2.3\\
363&360&1500&F2&2.2\\
439&360&1600&F1&2\\
449&450&1800&F1&1.8\\
602&240&2040&F2&1.7\\
603&320&1200&F2&1.7\\
708&1300&1500&F2&2.2\\
743&420&1500&F1&1.9\\
781&520&1200&F2&2.7\\
1002&300&1200&F2&2.0\\
1134&360&1320&F3&2.4\\
1194&300&1200&F3&1.9\\
1308&360&1800&F1&2.3\\
1379&360&1200&F2&1.9\\
\hline
\end{tabular}
\end{table}

\clearpage

\begin{table}
\caption{Derived properties of the sample galaxies.}
\begin{tabular}{lclllll}
\hline
\hline
Mkn &log $f_{H\alpha}$ & log $L_{H\alpha}$& EW(H$\alpha$)& n/g&$r_{e, H\alpha}$&$r_{e, H\alpha}/r_{e, R}$ \\
&(erg  s$^{-1}$ cm $^{-2}$)&erg  s$^{-1}$&\AA&& (kpc)&\\
(1)&(2)&(3)&(4)&(5)&(6)&(7)\\
\tableline
14& -12.40&40.96&66&1.31&0.55&0.37\\
213&-12.29&41.21&21&1.94&0.98&0.43\\
363&-12.29&41.02&53.2&0.59&1.17&0.64\\
439&-11.65&40.74&52&1.38&0.61&0.92\\
449&-12.17&40.28&35&1.63&0.23&0.17\\
602&-12.04&41.20&45&--&0.939&0.31\\
603&-11.39&41.79&150&1.17&0.93&0.65\\
708&-12.14&40.79&53&1.89&1.25&0.255\\
743&-12.04&40.17&101&1.15&0.19&0.49\\
781&-12.27&41.08&21&--&3.52&0.60\\
1002&-11.95&41.36&76&1.50&0.643&0.256\\
1134&-12.35&41.44&60&1.10&1.87&0.748\\
1194&-11.97&41.69&42&0.80&1.42&0.57\\
1308&-12.07&40.16&78&1.21&0.21&0.39\\
1379&-11.54&41.66&79&1.47&1.98&0.83\\
\tableline
\end{tabular}
\vskip 5mm
\tablecomments
{
Col. (1) : The Mkn number, Col. (2) : The total H$\alpha$+[NII]
flux of the galaxy corrected for galactic absorption using Eqn. \ref{ext},
Col. (3): The corrected
luminosity of the galaxy in logarithmic units, Col. (4) : The global equivalent
width in \AA, Col. (5) : The ratio of the nuclear (central 1 kpc) to the global EW,
Col. (6) : The half light radius in H$\alpha$, Col. (7) : The ratio of the half
light radius in H$\alpha$ to the half light radius in $R$ band.
}
\end{table}
\clearpage
\begin{table}
\caption{Average properties of our sample and LH96 sample}
\begin{tabular}{lll}
\hline
\hline
Parameter&this work& LH96\\
\hline
$S_{60 \mu m}$(Jy)&0.25-12.6&$>$5\\
$S_{60 \mu m}/S_{100 \mu m}$&0.23-0.83&$>$0.5\\
inclination&no limit&edge-on \\
log L$_{H\alpha}$&41.06&41.23\\
log L$_{FIR}$&43.25&44.22\\
log$r_{e,H\alpha}$(pc)&2.88&3.15\\
log($r_{e,H\alpha}/r_{e,R}$)&-0.34&-0.22\\
\hline
\end{tabular}
\end{table}
\clearpage
\section{References}
\begin{enumerate}
\item Burstein D., Heiles C., 1984, ApJS 54, 33
\item Bushouse H.A., 1986, AJ 91, 255
\item Chitre A.A., 1999, PhD thesis, Gujarat University
\item Chitre A., Joshi U.C., 1999, A\&AS 139, 105 (Paper I)
\item Chitre A., Joshi U.C., Ganesh S., 1999, A\&A 352, 112
\item Garc\'{i}a-Barreto, J.A., Franco, J., Carrillo, R., et al., 1996, Rev. Mex. Astron. Astrofis. 32, 89
\item Gonzalez Delgado, R.M., Perez, E., 1997a, ApJS 108, 155
\item Gonzalez Delgado, R.M., Perez, E., 1997b, ApJS 108, 199
\item Huchra J.P., 1977, ApJ 217, 926
\item Kennicutt R.C., 1983, ApJ 272, 54
\item Kennicutt R.C., JR., Kent S.M., 1983, AJ 88, 1094
\item Lehnert M.D., Heckman T.M., 1996, ApJ 472, 546
\item Lonsdale C.J., Helou G., Good J., et al., 1985, Catalogued Galaxies and Quasars in the IRAS Survey (Washington: GPO)
\item Macchetto F., Pastoriza M., Caon N., et al., 1996, A\&AS 120, 463
\item Markarian, B.E., Lipovetskii, V.A., Stepanian D.A., 1979, Astrofizika 15, 549
\item Mazzarella, J.M., Boroson, T.A., 1993, ApJS 85, 27
\item Pogge, R.W., 1989, ApJS 71, 433
\item Pogge, R.W., Eskridge, P.B., 1993, AJ 106, 1405
\item Romanishin W., 1990, AJ 100, 373
\item Usui T., Saito M., Tomita A., 1998, AJ 116, 2166
\item Whitford A., 1958, AJ 63, 201
\end{enumerate}

\section{List of figures}
\begin{enumerate}
\item Normalized response curves of the narrow band filters.

\item $R$ band continuum isophotal contours superimposed on the H$\alpha$ + 
[NII] emission line gray scale images. The gray scale has been chosen for best
contrast. North is at the top and East is to the left. The $R$ band contours are
plotted on the magnitude scale at an interval of 0.5$^m$ with the peak contours 
at 16.5 mag arcsec$^{-2}$ for Mkn 213, Mkn 603 and Mkn 1379; 17.0 mag arcsec$^{-2
}$ for Mkn 439, Mkn 602 and Mkn 1194; 17.5 mag arcsec$^{-2}$ for Mkn 449, Mkn
708 and Mkn 781; 18.0 mag arcsec$^{-2}$ for Mkn 14, Mkn 363, Mkn 743, Mkn 1002 and
 Mkn 1308 and 18.5 mag arcsec$^{-2}$ for Mkn 1134.

\item H$\alpha$ equivalent widths (present study) vs. B-V colours (Chitre 1999)

\item The radial variation of the logarithm of the H$\alpha$ flux (erg  s$^{-1}$ cm $^{-2}$) and the EW (\AA)

\item The absolute B magnitude as a function of the half-light radius in H$\alpha$

\item log($L_{IR}/L_{H\alpha}$) vs r$_{e,H\alpha}$/r$_{e,r}$ for our sample compared with two other samples.
\end{enumerate}

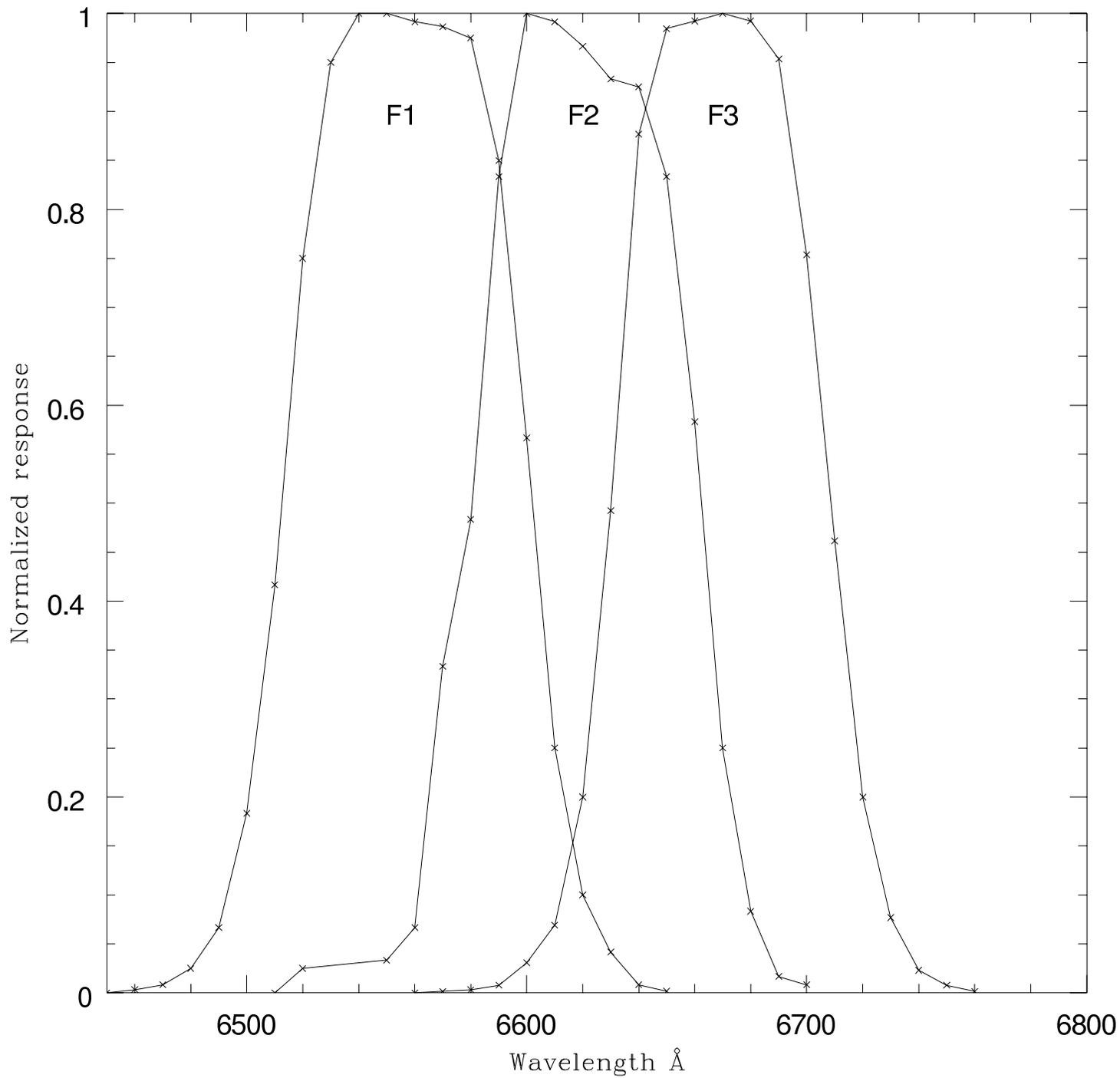
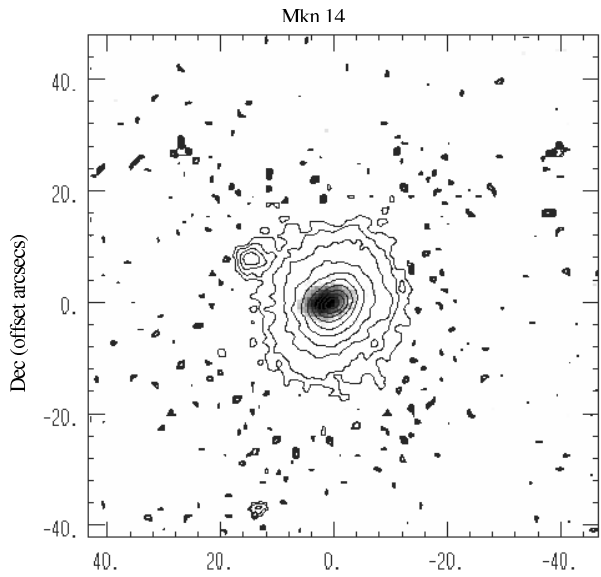
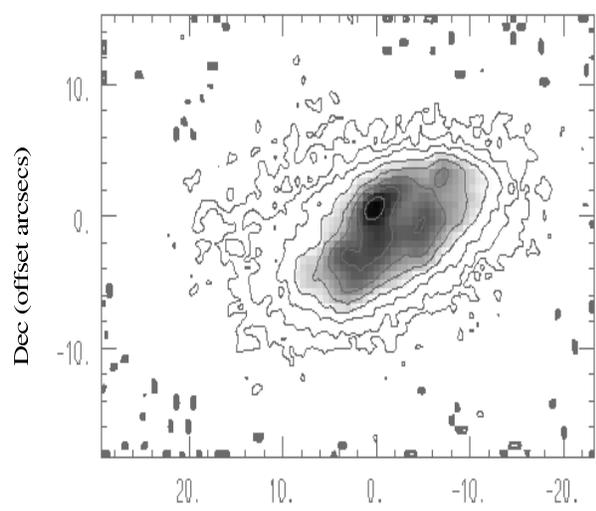
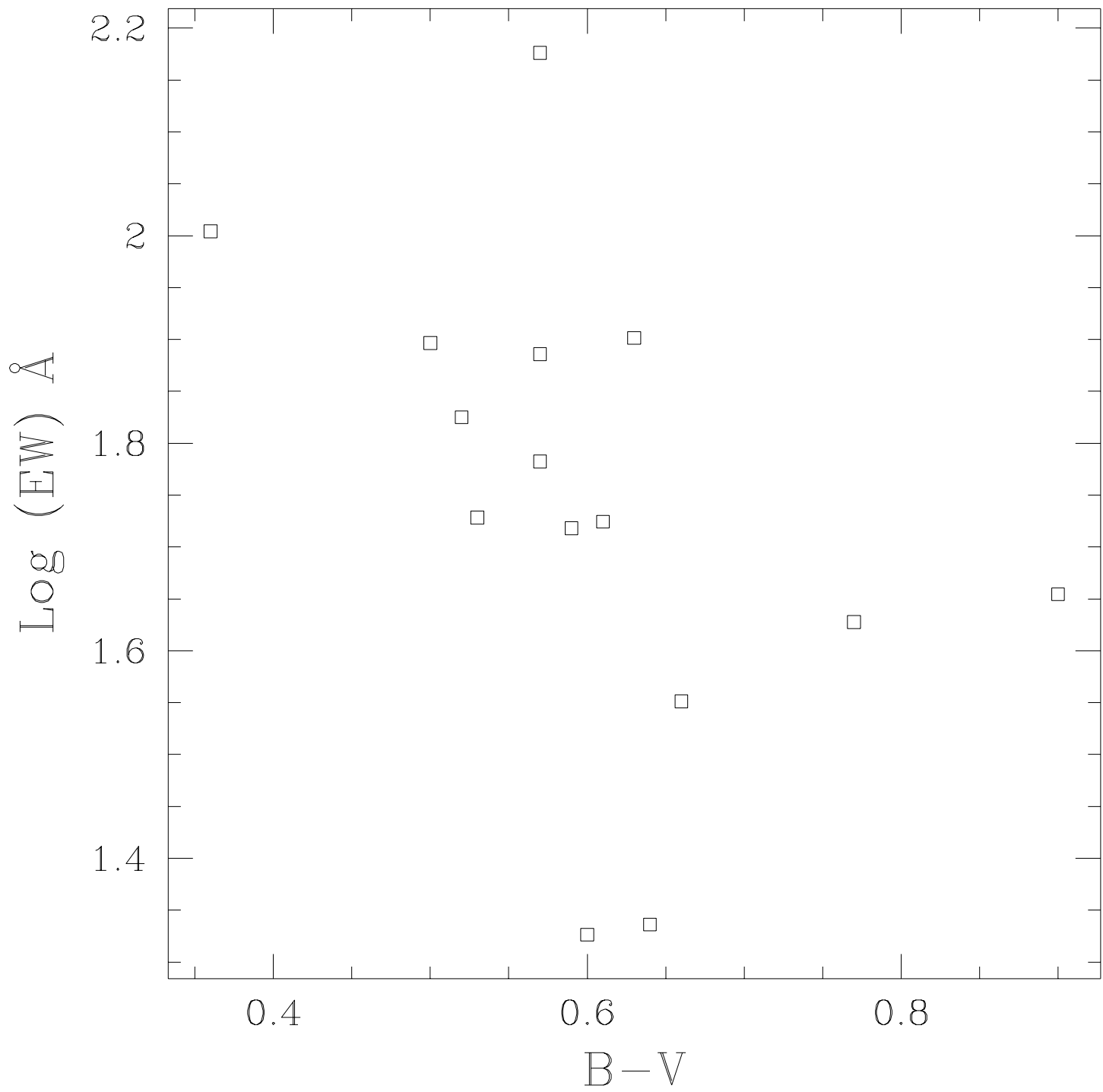
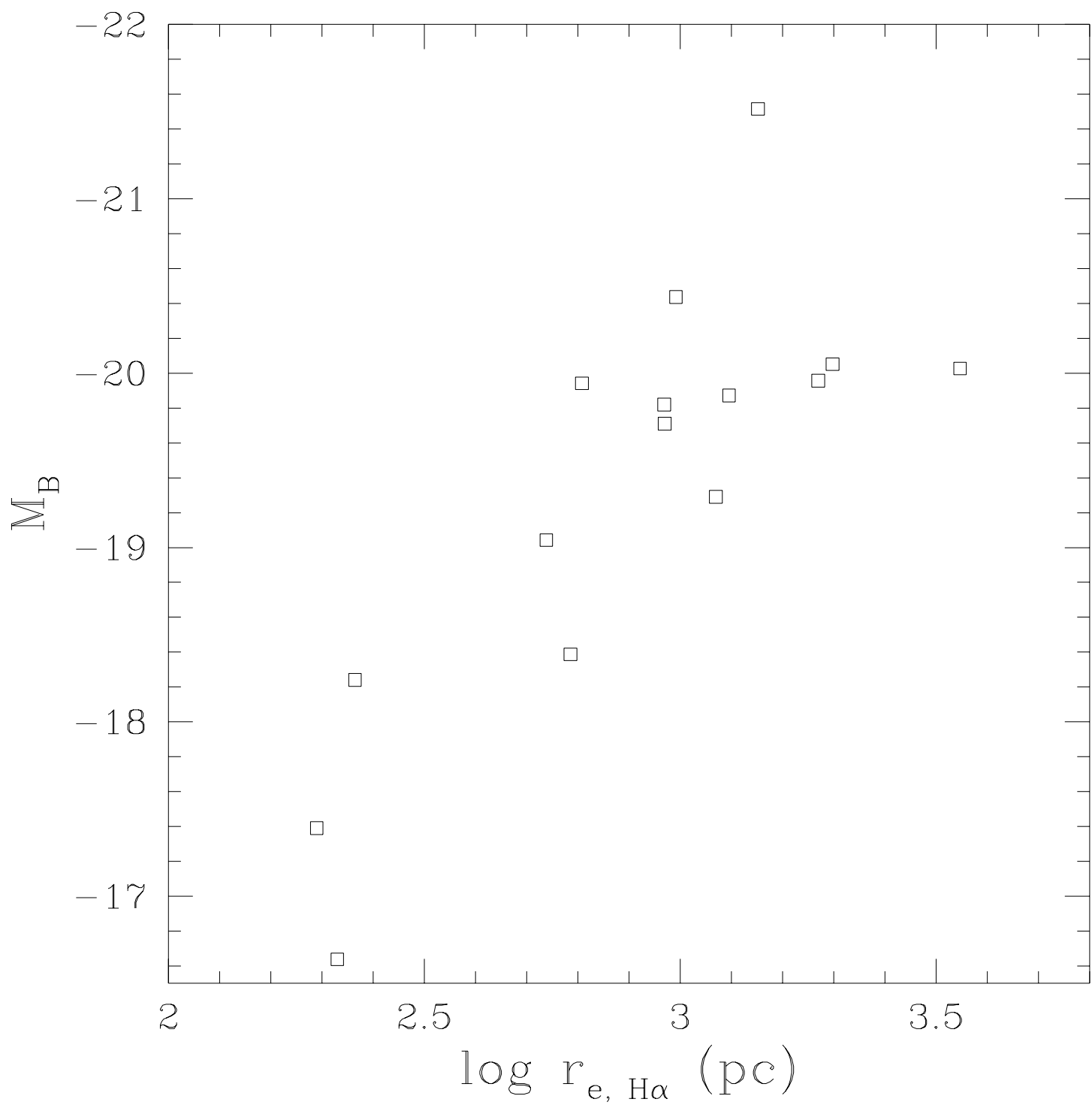
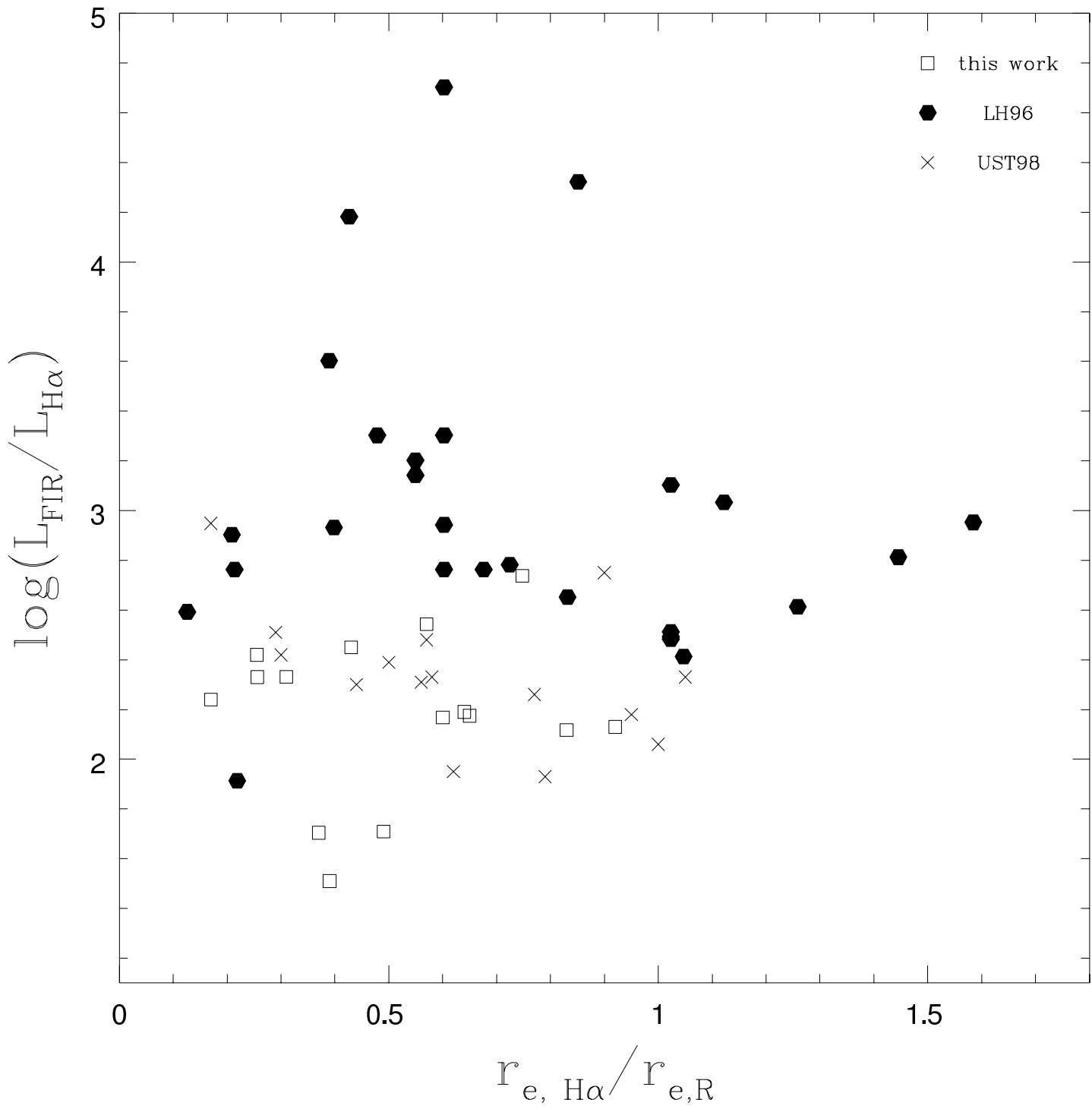
\end{document}